\documentclass[a4paper]{article}
\usepackage[dvips]{graphicx}
\usepackage[margin=3cm]{geometry}
\newtheorem{definition}{Definition}
\newtheorem{corollary}[definition]{Corollary}
\newtheorem{theorem}[definition]{Theorem}
\newtheorem{lemma}[definition]{Lemma}
\newtheorem{conjecture}[definition]{Conjecture}

\begin{document}
\title{Families of line-graphs and their quantization}
\author{Prot Pako\'nski$^1$, Gregor Tanner$^2$ and Karol \.Zyczkowski$^3$ \\
  \small $^1$Uniwersytet Jagiello\'nski,
         Instytut Fizyki im. M.~Smoluchowskiego, \\
  \small ul. Reymonta 4, 30--059 Krak\'ow, Poland \\
  \small $^2$School of Mathematical Sciences,
         Division of Theoretical Mechanics, \\
  \small University of Nottingham, University Park, Nottingham NG7 2RD, UK \\
  \small $^3$Centrum Fizyki Teoretycznej, Polska Akademia Nauk, \\
  \small Al. Lotnik\'ow 32/44, 02--668 Warszawa, Poland \\
  \small e-mail addresses: \emph{pakonski@if.uj.edu.pl},
         \emph{gregor.tanner@nottingham.ac.uk}, \emph{karol@cft.edu.pl} }
\maketitle
\begin{abstract}
  Any directed graph $G$ with $N$ vertices and $J$ edges has an
  associated line-graph $L(G)$ where the $J$ edges form the vertices
  of $L(G)$. We show that the non-zero eigenvalues of the adjacency
  matrices are the same for all graphs of such a family $L^n(G)$. We
  give necessary and sufficient conditions for a line-graph to be
  quantisable and demonstrate that the spectra of associated quantum
  propagators follow the predictions of random matrices under very
  general conditions. Line-graphs may therefore serve as models to study
  the semiclassical limit (of large matrix size) of a quantum dynamics
  on graphs with fixed classical behaviour. 
\end{abstract}

\section{Introduction}
Spectra of quantum graphs display in general universal statistics
characteristic for ensembles of random unitary matrices. This
observation by Kottos and Smilansky \cite{Ko97, Ko99} has led to a
variety of studies in this direction \cite{Be99, Sc99, Sc00, Ko00,
Ba00, Be00, Ba01, Be02, Ta02}. It has became clear that the
quantisation scheme of Kottos and Smilansky can be considerably
generalised to be applicable also for directed graphs (digraphs)
\cite{Ta00, Ta01, Pa01}. One of the main points of interest is to
understand under which circumstances the quantisation of a graph
generates a spectrum which follows random matrix theory (RMT) and when
to expect deviations thereof. General statements can, however, only be
made in the limit of large matrices and we thus face the problem of
constructing larger and larger graphs representing the same classical
dynamics at least in the limit of infinite network size. We will offer
a simple and straightforward way to define such families of graphs in
this paper.

We thereby consider families of graphs generated from an arbitrary
initial graph by using the concept of \emph{line-graphs} \cite{Cv84,
BJ01} (also called \emph{edge-graphs}). Consider any initial directed
graph $G$ with $N$ vertices and $J$ bonds (edges). The line-graph
$L(G)$ obtained from $G$ consists of $J$ vertices which are the edges
of its ancestor $G$. Iterating this procedure we construct an infinite
family of digraphs $L^n(G)$ with in general increasing number of
vertices. We will show that all graphs in a given family defined in
this way have the same topological and metric properties.
In particular, the sets of periodic orbits and the non-zero
eigenvalues of the adjacency and transition matrices are identical for
digraphs of such a family. We will give necessary and sufficient
conditions for a line-graph to be quantisable. 

Line-graph families thus form a set of graphs whose size increases
with $n$ but whose `classical' dynamics is fixed. The semiclassical
limit of the system is then obtained by increasing the index $n$. The
entire family of graphs, corresponding to the same classical dynamics,
is uniquely determined by a given initial graph. This approach to the
semiclassical limit for quantum graphs offers an alternative to the
previous method based on transition matrices representing Markov
chains associated with certain piecewise linear 1D dynamical systems
\cite{Pa01}. 

Our paper is organized as follows. In section~\ref{seceg} we recall
the definition of a line-graph and present an example of a family of
digraphs. Sections~\ref{secpg} and~\ref{seclg} contain the main
results of this work: a proof that all graphs belonging to a given
family of line-graphs represent the same dynamics and conditions for
the quantisability of line-graphs. Section~\ref{secex} is devoted to
examples of quantisable line-graph families. We analyze in particular
the statistical properties of the spectra of the unitary matrices
obtained when quantising the graph. Concluding remarks are presented
in section~\ref{secco}.

\section{Line-graphs - definitions and basic properties} \label{seceg}
Consider a directed graph $G$ with $N$ vertices and $J$ edges (called
also bonds or arcs). We denote the set of vertices $V(G) = \{v_1,
\ldots, v_N\}$ and the set of edges by $E(G) = \bigl\{ (v_i v_j):
G \mbox{ has an edge leading from $v_i$ to $v_j$} \bigr\}$. To
simplify the notation, we will only consider graphs with at most one edge
going from a vertex $v_i$ to a vertex $v_j$. All the results in this
paper apply, however, also for directed multi-graphs $G$, i.e. for
graphs with two or more edges connecting two vertices in the same
direction. We will use the ordered pair $(ij)$ to represent a directed
edge. A digraph $G$ may have loops, i.e. edges starting and ending at
the same vertex. A \emph{line-graph} $L(G)$ is constructed from
a graph $G$ by considering the edges as vertices, that is,
\begin{equation}
  V\bigl( L(G) \bigl) = E(G),
\end{equation}
and vertices in $L(G)$ are adjacent if the edges in $G$ are. It is
clear from the definition that $L(G)$ does not have multi-edges even
if $G$ does; one obtains
\begin{equation}
  E\bigl( L(G) \bigr) = \Bigl\{ \bigl( (ij),(jk) \bigr):
    (ij) \in E(G), (jk) \in E(G) \Bigr\} .
\end{equation}

We will be interested in families of digraphs obtained from $G$ by
iterating the line-graph procedure. The \emph{$n$-th generation
line-graph} $L^n(G)$ of $G$ is thereby defined as $L^n(G) = L\bigl(
L^{n-1}(G) \bigr)$. We will call the graph $L^{n-1}(G)$ the
\emph{ancestor} of the line-graph $L^n(G)$ and $G$ the \emph{initial
graph} of the family.

In what follows, we will need the set of vertices which can be reached
from a vertex $v_i$ in $n$ steps. We define the \emph{$n$-step
out-neighbourhood} of $v_i$ as
\begin{equation}
  N_+^{(n)} (v_i) = \bigl\{ v_j \in V(G): \mbox{ $v_j$ can be reached
    from $v_i$ in $n$ steps} \bigr\} ;
\end{equation}
equivalently, we define the \emph{$n$-step in-neighbourhood} of $v_i$
as
\begin{equation}
  N_-^{(n)} (v_i) = \bigl\{ v_j \in V(G): \mbox{ $v_i$ can be reached
    from $v_j$ in $n$ steps} \bigr\} .
\end{equation}
The cardinality (i.e. the number of elements) of $N_{\pm}^{(1)} (v_i)$
is often called the \emph{out/in-degree}, $d^{\pm}(v_i)$, of $v_i$
corresponding to the number of vertices adjacent to $v_i$ with respect
to outgoing or incoming edges.

The topology of a digraph $G$ is most conveniently described in terms
of its \emph{connectivity} or \emph{adjacency matrix} $A^{G}$ of size
$N$ with
\begin{equation}
  A^G_{ij} = \left\{ \begin{array}{lr} 1 & (ij)\in E(G) \\ 0 &
    (ij)\not\in E(G) \end{array} \right. \qquad i,j=1\ldots N \ .
\end{equation}
The degree of a vertex $v_i$ is then given as
\begin{equation}
  d^+(v_i) = \sum_{j=1}^{N} A^G_{ij} \qquad {\mathrm{and}} \qquad
  d^-(v_j) = \sum_{i=1}^{N} A^G_{ij} \, .
\end{equation}
The adjacency matrix of the line-graph $L(G)$ of $G$ may be expressed
as
\begin{equation}
  A^{L(G)}_{ij,kl} = A^G_{ij} \, \delta_{jk} \, A^G_{kl} \, .
\end{equation}
In fact if we define $A^{L(G)}$ as the adjacency matrix of dimension
$J$ including only the relevant index pairs $(ij), (kl) \in E(G)$ then
$A^{L(G)}_{ij,kl} = \delta_{jk}$.

A stochastic Markov - process on the graph $G$ is defined in terms of
a transition matrix $T^G$ with $T^G_{ij} \geq 0$ representing the
probability of going from vertices $i$ to $j$. We demand that $T^G$
has the same zero-pattern as $A^G$, that is $A^G_{ij}\ne 0$ iff
$T^G_{ij} \ne 0$ for all $i,j =1,\ldots N$; furthermore stochasticity
of $T^G$ implies that $\sum_j T^G_{ij}=1$. We define the transition
matrix $T^{L(G)}$ of the stochastic process induced by $T^{G}$ on the
line-graph of $G$ by
\begin{equation}
  T^{L(G)}_{ij,kl} = A^G_{ij} \, \delta_{jk} \, T^G_{kl} \ .
\end{equation}
It is obvious from the definition that $T^{L(G)}$ is a stochastic
matrix which has the same zero-pattern as $A^{L(G)}$.
\begin{figure}[hbt]
  \begin{center} \includegraphics{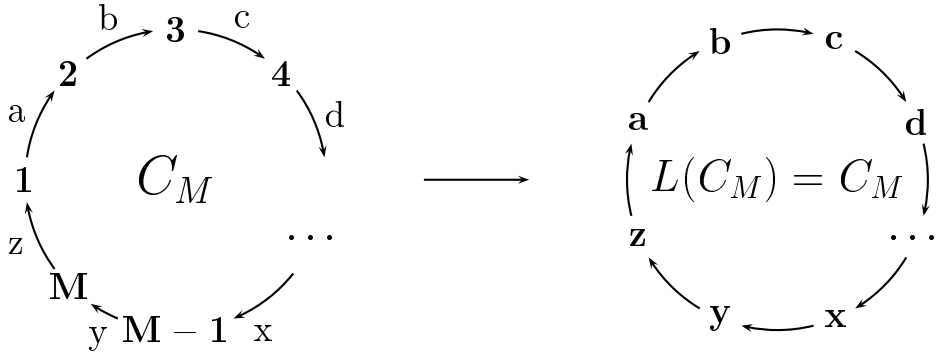} \end{center}
  \caption{Directed cycle digraph $C_M$, its line-graph $L(C_M)$ is
    isomorphic to it.}
  \label{ficyc}
\end{figure}

Before moving on to general results on line-graph families, we will
discuss a particular example to see how this construction works.
Consider first a directed cycle digraph $C_M$ (see Fig.~\ref{ficyc})
consisting of $M$ vertices connected by $M$ bonds. Such a graph is
strongly connected, that is, there exists at least one directed path 
leading from
a vertex $v_i$ to $v_j$ for all $i,j = 1, \ldots, M$. The line-graph
$L(C_M)$ is isomorphic to $C_M$ (see Fig.~\ref{ficyc}), so all
cycles $C_M$ are fixed points of the line-graph construction, $L(C_M)
= C_M$.
\begin{figure}[hbt]
  \begin{center} \includegraphics{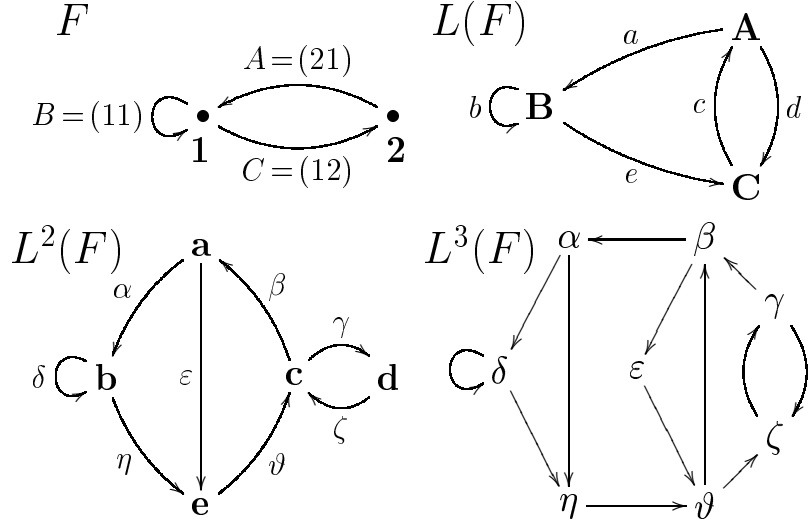} \end{center}
  \caption{Fibonacci family of line-graphs; the initial graph $F$, and
    next three members of the line-graph family consisting of 2, 3, 5
    and 8 vertices, respectively, are shown.}
  \label{fifib}
\end{figure}

Let us discuss next a family of digraphs generated by the initial
graph $F$ defined as
\begin{equation}
  V(F) = \{v_1, v_2\} \ , \qquad E(F) = \Bigl\{ (v_1v_1), (v_1v_2),
    (v_2v_1) \Bigl\} = \Bigl\{ (11), (12), (21) \Bigr\} \ .
\end{equation}
Fig.~\ref{fifib} shows the first four graphs of this family. Their
adjacency matrices are
\begin{displaymath}
  C^{F} = \left( \begin{array}{cc}
    1 & 1 \\ 1 & 0 \end{array} \right),
    \quad C^{L(F)} = \left( \begin{array}{ccc}
    1 & 1 & . \\ . & . & 1 \\ 1 & 1 & . \end{array} \right),
    \quad C^{L^2(F)} = \left( \begin{array}{ccccc}
    1 & 1 & . & . & . \\ . & . & 1 & . & . \\ . & . & . & 1 & 1 \\
    1 & 1 & . & . & . \\ . & . & 1 & . & . \end{array} \right),
\end{displaymath}
\begin{equation}
  C^{L^3(F)} = \left( \begin{array}{cccccccc}
    1 & 1 & . & . & . & . & . & . \\
    . & . & 1 & . & . & . & . & . \\
    . & . & . & 1 & 1 & . & . & . \\
    . & . & . & . & . & 1 & 1 & . \\
    . & . & . & . & . & . & . & 1 \\
    1 & 1 & . & . & . & . & . & . \\
    . & . & 1 & . & . & . & . & . \\
    . & . & . & 1 & 1 & . & . & . \end{array} \right);
\end{equation}
the dots represent here entries being equal to zero. To introduce a
stochastic process on $F$ we may choose equal probabilities of staying
at vertex 1 and of going from 1 to 2. This corresponds to the
transition matrix
\begin{equation}
  T^{F} = \frac{1}{2}\left( \begin{array}{cc}
    1 & 1 \\ 2 & 0 \end{array} \right).
\end{equation}
The transition matrices $T^{L^n(F)}$ can be obtained from
$A^{L^n(F)}$ by replacing 1's by $\frac{1}{2}$'s in all rows in
which there are two entries equal to unity. The resulting
matrices are stochastic. Let $N_G$ denote the number of
vertices of a digraph $G$. Then the numbers of vertices of
the digraphs $L^n(F)$ fulfill the Fibonacci relation
\begin{equation}
  N_{L^n(F)} = N_{L^{n-1}(F)} + N_{L^{n-2}(F)} \ ,
\end{equation}
for $n>1$ with $N_{F}=2$ and $N_{L(F)}=3$.
 
\section{Line-graph families $L^n(G)$ - stochastic dynamics} \label{secpg}
We shall start this section by stating basic properties of the
line-digraph construction. If $G$ is a strongly connected digraph not
isomorphic to a cycle, then the number of its bonds is larger than
the number of its vertices, so
\begin{equation}
  N_{L(G)} > N_G \, .
\end{equation}
Observe that $L(G)$ is also a strongly connected digraph different
from a cycle. The above statements allow us to draw an important
conclusion:
\begin{corollary}
  For any strongly connected digraph $G$, not isomorphic to a cycle,
  its line-graph family $L^n(G)$ contains infinite number of different
  digraphs and
  \begin{equation}
    \lim_{n \rightarrow \infty} N_{L^n(G)} = \infty \, .
  \end{equation}
\end{corollary}
In the following we analyze topological and dynamical properties of
line-graph families $L^n(G)$ with associate stochastic Markov
processes. We start by introducing periodic orbits on a digraph.
\begin{definition}
  A sequence of $p$ vertices $\gamma=v_{i_1}v_{i_2}\ldots v_{i_p}$
  such, that $v_{i_j}\in V(G),\ j=1\ldots p$ and $(v_{i_j}
  v_{i_{j+1}}) \in E(G),\ j=1\ldots p-1,\ (v_{i_p} v_{i_1}) \in E(G)$
  is called a periodic orbit of period $p$ on the digraph $G$. The set
  of periodic orbits on $G$ is denoted by $PO(G)$.
\end{definition}
A periodic orbit is called \emph{primitive}, if it is not a repetition
of another periodic orbit. It is obvious from the definition of a
line-graph that there is a one-to-one relation between periodic orbits
of $G$ and $L(G)$, that is, $\gamma = v_{i_1} v_{i_2} \ldots v_{i_p}
\in PO(G)$ iff $\eta = (v_{i_1} v_{i_2}) (v_{i_2} v_{i_3}) \ldots
(v_{i_{p-1}} v_{i_p}) (v_{i_p} v_{i_1}) \in PO(L(G))$. The set of
periodic orbits $PO(L^n(G))$ is thus isomorphic to $PO(G)$ and the map
\begin{equation}
  \Theta: PO(G) \rightarrow PO(L(G))
\end{equation}
between periodic orbits of $G$ and $L(G)$ is bijective and conserves
the period of the orbit. This implies that the topological entropy
measuring the exponential of growth of the number of periodic orbits
with their period $p$ is the same for all generations of the
line-graph family. The four graphs presented in Fig.~\ref{fifib} may
serve as an example. All the graphs $L^n(F)$ have only one primitive
orbit of periods 1, 2, 3 and 4.

Next, we define the \emph{stability factor} or \emph{amplitude} of a
periodic orbit, $\gamma=v_{i_1}v_{i_2}\ldots v_{i_p}\in PO(G)$ of a
graph $G$ with associated stochastic process $T^G$ as
\begin{equation}
  a^G_\gamma = T^G_{i_1i_2} \cdot T^G_{i_2i_3}
    \cdot \ldots \cdot T^G_{i_pi_1} \, .
\end{equation}
The amplitude $a^G_\gamma$ is the probability of staying on the orbit
$\gamma$ after $p$ iterations of the stochastic process, where $p$ is
the period of $\gamma$. One obtains for the stability factor of
periodic orbits on the line-graph
\begin{equation}
  a^{L(G)}_{\Theta(\gamma)} = T^{L(G)}_{i_1i_2,i_2i_3}
    \cdot T^{L(G)}_{i_2i_3,i_3i_4} \cdot \ldots \cdot
    T^{L(G)}_{i_pi_1,i_1i_2} = A^G_{i_1i_2} T^G_{i_2i_3}
    \cdot A^G_{i_2i_3} T^G_{i_3i_4} \cdot \ldots \cdot
    A^G_{i_pi_1} T^G_{i_1i_2} = a^G_{\gamma},
\end{equation}
and the last identity follows from
\begin{equation} \label{eqpat}
  A^G_{ij} \cdot T^G_{ij} = T^G_{ij} \, .
\end{equation}
We thus obtain that the mapping $\Theta$ leaves the stability factors
of periodic orbits invariant, that is,
\begin{equation}
  a^{L(G)}_{\Theta(\gamma)} = a^G_\gamma \, .
\end{equation}
 
The observations made above on the invariance of topological and
dynamical properties under the line-graph construction can be put in a
more general form. The topological entropy of a graph may be
determined by the logarithm of the largest modulus of eigenvalue of
the adjacency matrix of the graph. Denoting the characteristic
polynomial of the adjacency matrix by
\begin{equation} \label{eqcpa}
  P^G(\lambda) = \det(A^G-\lambda{\mathbf{1}})
\end{equation}
one obtains:
\begin{theorem} \label{thsam}
  The spectrum of the adjacency matrix of the line-graph, $A^{L(G)}$,
  is identical to the spectrum of $A^G$ and an appropriate number of
  eigenvalues equal to zero, that is,
  \begin{equation}
    P^{L(G)}(\lambda) = P^G(\lambda) \cdot (-\lambda)^{N_{L(G)}-N_G} \ .
  \end{equation}
\end{theorem}
Proof: We start by the following lemma.
\begin{lemma} \label{letam}
  The traces of powers of the adjacency matrix of a line-graph,
  $A^{L(G)}$, are equal to the trace of the same power of $A^G$, that
  is,
  \begin{equation}
    {\mathrm{Tr}}\left(A^{L(G)}\right)^n =
      {\mathrm{Tr}}\left(A^G\right)^n \qquad {\mathrm{for\ all\ }} n.
  \end{equation}
\end{lemma}
Since all entries of any adjacency matrix are equal to 0 or to 1 we
have $A^G_{ij} \cdot A^G_{ij} = A^G_{ij}$. One thus obtains 
\begin{displaymath}
  \mbox{Tr}\left(A^{L(G)}\right)^n = \sum_{(i_1j_1)(i_2j_2)
    \ldots(i_nj_n)\in E(G)} A^{L(G)}_{i_1j_1,i_2j_2}
    A^{L(G)}_{i_2j_2,i_3j_3} \ldots A^{L(G)}_{i_nj_n,i_1j_1} =
\end{displaymath}
\begin{displaymath}
  = \sum_{i_1\ldots i_nj_1\ldots j_n\in V(G)}
    \left(A^G_{i_1j_1}\delta_{j_1i_2}A^G_{i_2j_2}\right)
    \left(A^G_{i_2j_2}\delta_{j_2i_3}A^G_{i_3j_3}\right)\ldots
    \left(A^G_{i_nj_n}\delta_{j_ni_1}A^G_{i_1j_1}\right) =
\end{displaymath}
\begin{equation}
  = \sum_{i_1\ldots i_n\in V(G)}
    A^G_{i_1i_2}A^G_{i_2i_3}A^G_{i_3i_4}\ldots
    A^G_{i_ni_1} = \mbox{Tr}\left(A^G \right)^n .
\end{equation}
Let $\tau_k$ denote the coefficients of the characteristic polynomial
of $A^G$ in the descending order
\begin{equation}
  P^G(\lambda) = (-\lambda)^{N_G}-\tau_1(-\lambda)^{N_G-1}
  +\tau_2(-\lambda)^{N_G-2}-\ldots(-1)^{N_G}\tau_{N_G} \ .
\end{equation}
By means of the Newton formulas the coefficients $\tau_k$ may be
expressed in terms of the traces $D_n := \mbox{Tr} \left( A^G
\right)^n$ as \cite{Mo64}
\begin{equation}
  \tau_k = \frac{1}{k!}\left|\begin{array}{ccccc}
        D_1  &    1    &     0   & \ldots &    0   \\
        D_2  &   D_1   &     2   & \ldots &    0   \\
        D_3  &   D_2   &    D_1  & \ldots &    0   \\
          \vdots &  \vdots &  \vdots & \vdots & \vdots \\
        D_k  & D_{k-1} & D_{k-2} & \ldots &   D_1
       \end{array}\right| .
\end{equation}
Lemma~\ref{letam} shows that the first $N_G$ coefficients of the
polynomial $P^{L(G)}$ in front of the largest powers of $\lambda$ are
equal to those of $P^G$. The rest of the coefficients of $P^{L(G)}$
vanish, the characteristic polynomials of $A^{L(G)}$ and $A^G$
differ thus only by a factor $(-\lambda)^{N_{L(G)}-N_G}$; this completes
the proof of the theorem~\ref{thsam}.

A relation similar to (\ref{eqcpa}) holds for the characteristic
polynomial of $T^G$
\begin{equation}
  R^G(\lambda) = \det(T^G-\lambda{\mathbf{1}}) \ .
\end{equation}
One obtains:
\begin{theorem} \label{thetm}
  The spectrum of the transition matrix of a line-graph, $T^{L(G)}$
  consists of the spectrum of $T^G$ and an appropriate number of
  eigenvalues equal to zero, so
  \begin{equation}
    R^{L(G)}(\lambda) = R^G(\lambda) \cdot (-\lambda)^{N_{L(G)}-N_G} \ .
  \end{equation}
\end{theorem}
Proof is analogous to this of the theorem~\ref{thsam}, since a lemma
equivalent to the lemma~\ref{letam} holds:
\begin{lemma}
  Traces of any power of the transition matrix of a line-graph
  $T^{L(G)}$ are equal to the trace of the same power of $T^G$, that
  is
  \begin{equation}
    {\mathrm{Tr}}\left(T^{L(G)}\right)^n = {\mathrm{Tr}}\left(T^G\right)^n .
  \end{equation}
\end{lemma}
The derivation follows the arguments in the proof of lemma~\ref{letam}
using the property~(\ref{eqpat}) instead.
 
Theorem~\ref{thetm} demonstrates that the stochastic dynamics on the
line-graph $L(G)$ is equivalent to the original Markov process on $G$.
We would therefore expect that dynamical quantities like the metric
entropy of the stochastic process are invariant under the line-graph
iteration as well. The metric entropy depends on the choice of the
invariant measure, so we need to consider invariant measures first.
The action of $T^G$ on left vectors represents the evolution of
measures. One obtains
\begin{lemma} \label{lelvt}
  If $\rho_i$ is a left eigenvector of $T^G$ corresponding to the
  eigenvalue $\lambda$, then $(\rho_i T^G_{ij})$ is the left
  eigenvector of $T^{L(G)}$ to the same eigenvalue.
\end{lemma}
Proof: Let us calculate
\begin{equation}
  \sum_{(ij)\in E(G)} (\rho_i T^G_{ij}) T^{L(G)}_{ij,kl}
    = \sum_{i,j\in V(G)} \rho_i T^G_{ij} A^G_{ij} \delta_{jk}
    T^G_{kl} = \sum_{i\in V(G)} \rho_i T^G_{ik} T^G_{kl}
    = \lambda (\rho_k T^G_{kl}) \ ,
\end{equation}
where we have used~(\ref{eqpat}) and the fact that $\rho_i$ is the
left eigenvector of $T^G$,
\begin{equation}
  \sum_{i\in V(G)}\rho_i T^G_{ij} = \lambda\rho_j \ .
\end{equation}

The invariant measures of a Markov chain on $G$ is given by the left
eigenvectors of $T^G$ with eigenvalue 1. According to
Lemma~\ref{lelvt} each invariant measure of $T^G$ defines the
corresponding invariant measure of $T^{L(G)}$. Assuming that $\rho^G$
is an invariant measure of $T^G$, the metric entropy of the
corresponding Markov process \cite{Ka95} reads
\begin{equation}
  H^G_{\rm metric} = - \sum_{i\in V(G)} \rho^G_i
    \sum_{j\in V(G)} T^G_{ij} \ln T^G_{ij} \ .
\end{equation}
The metric entropy of the Markov process on $L(G)$ with respect to the
corresponding invariant measure $\rho^{L(G)}_{ij}=\rho^G_i T^G_{ij}$
is then given as (see Lemma~\ref{lelvt})
\begin{displaymath}
  H^{L(G)}_{\rm metric} = - \!\!\!\!\sum_{(ij)\in E(G)}\!\!\!\!
    \rho^{L(G)}_{ij} \!\!\!\!\sum_{(kl)\in E(G)}\!\!\!\!
    T^{L(G)}_{ij,kl} \ln T^{L(G)}_{ij,kl} = -
    \!\!\!\!\sum_{ijkl\in V(G)}\!\!\!\! \rho^G_i T^G_{ij}
    A^G_{ij} \delta_{jk} T^G_{kl}
    \ln A^G_{ij} \delta_{jk} T^G_{kl}
\end{displaymath}
\begin{equation}
  = - \sum_{ijl\in V(G)} \rho^G_i T^G_{ij}
    T^G_{jl} \left( \ln A^G_{ij}+\ln T^G_{jl}
    \right) = - \sum_{jl\in V(G)} \rho^G_j T^G_{jl}
    \ln T^G_{jl} \ .
\end{equation}
We thus find that the metric entropy of a stochastic process defined
by $T^{L(G)}$ based on the invariant measure $\rho^{L(G)}$ is indeed
identical to the metric entropy of a process $T^G$ based on the
invariant measure $\rho^G$, that is,
\begin{equation}
  H^{L(G)}_{\rm metric} = H^G_{\rm metric} \ .
\end{equation}
 
The results stated in this section show that the topological and
metric properties of the dynamics on a given graph $G$ and the
corresponding line-graph $L(G)$ are identical. In fact we have proven
by recurrence that all digraphs in the family $L^n(G)$ have the same
set of periodic orbits, the same non-vanishing spectrum of the
adjacency matrices $A^{L^n(G)}$ and of the transition matrices
$T^{L^n(G)}$, as well as the same topological and metric entropy.
 
\section{The quantisation of line-graph families} \label{seclg}

\subsection{Unitary propagation on graphs}
So far we have considered stochastic processes on digraphs defined by
a transition matrix $T^G$. Recently, Kottos and Smilansky \cite{Ko97}
proposed to study unitary propagation on graphs and to link the
spectral properties of the unitary dynamics to an underlying Markov
process on this graph. Generalising their approach we may consider the
following definition of quantising a Markov chain:
\begin{definition} \mbox{}
  \begin{enumerate}
    \renewcommand{\theenumi}{(\alph{enumi})}
    \renewcommand{\labelenumi}{\theenumi}
  \item
    A digraph $G$ is called \emph{quantisable} if there exists a
    unitary matrix $U^G$ with the same zero-pattern as the adjacency
    matrix $A^{G}$.
  \item
    A stochastic transition matrix $T^G$ is called \emph{quantisable}
    if there exists a unitary matrix $U^G$ such that
    \begin{equation} \label{eqqtc}
      T^G_{ij} = \left| U^G_{ij} \right|^2 \ .
    \end{equation}
  \end{enumerate}
\end{definition}
The matrix $U^G$ represents a one-step propagator, which describes 
unitary time evolution in a finite Hilbert space of dimension $N_G$.
Note that not all stochastic matrices $T^G$ can be quantised in the
sense described above. The stochastic matrices, for which a unitary
matrix exists fulfilling eqn. (\ref{eqqtc}) are called
\emph{unistochastic} \cite{Ma79}. The matrix $T^G$ in (\ref{eqqtc}) is
by construction bistochastic, that is, the sum over the matrix
elements in each row and column of $T^G$ equals $1$. However, for
$N_G>2$ bistochasticity is not a sufficient condition for
unistochasticity (see e.g. \cite{Ma79, Be01, Zy01, Pa01}), and it is
in general hard to decide whether a given bistochastic matrix is
unistochastic or not. Even necessary and sufficient conditions for 
the pattern of unitary matrices are not known, see \cite{Se02} for some 
necessary conditions.

On the other hand, the quantisation of a unistochastic matrix $T^G$ is
not unique. For every matrix $U^G$ fulfilling (\ref{eqqtc}), the set
of unitary matrices of the form
\begin{equation} \label{equse}
  \tilde{U}^G = D_1 U^G D_2 \ ,
\end{equation}
with $D_1$ and $D_2$ being diagonal unitary matrices, are also
quantisations of $T^G$. One can therefore introduce a $2N_G-1$
parameter family of unitary matrices corresponding to the same
classical stochastic process defined by $T^G$. By choosing the phases
in $D_1$ and $D_2$ randomly with respect to the uniform measure on the
interval $[0,2\pi)$ one can define an ensemble of unitary matrices
linked to the transition matrix $T^G$ as proposed in \cite{Ta01}, also
called a \emph{unitary stochastic ensemble (USE)} of $T^G$. The
transition matrix $T^G$ is stochastic and its largest eigenvalue
$\lambda_1$ is equal to unity \cite{Ma79}. It was conjectured in
\cite{Ta01} that the statistical properties of spectra of unitary
matrices in a given $USE$ after ensemble average are linked to the
\emph{spectral gap} $\Delta_{T^G} = 1 - |\lambda_2|$ of $T^G$, where
$\lambda_2$ is the subleading eigenvalue of $T^G$. The conjecture in
\cite{Ta01} implies in particular the following
\begin{conjecture} \label{cormt}
  Let $T(N)$ be a family of unistochastic transition matrices of
  dimension $N$; the corresponding unitary stochastic ensembles follow
  random matrix theory (RMT) in the limit $N \to \infty$ if the
  spectral gap is bounded from below, that is, if $\Delta_{T(N)} \ge c
  > 0$ in this limit.
\end{conjecture}
It has been shown in the last section, that the spectral gap remains
constant for stochastic processes generated by line-graph iterations.
The conjecture thus implies that unistochastic ensembles derived from
quantisable line-graph families $L^n(G)$ follow RMT in the limit
$n\to\infty$ (assuming $G$ is a strongly connected digraph not
isomorphic to a cycle) if the spectral gap of the Markov chain on
the initial graph $\Delta_{T(G)} > 0$. As mentioned above not all
stochastic processes on digraphs are quantisable in the sense above
and it is in general hard to decide whether a given bistochastic
transition matrix is unistochastic or not or even whether a given
graph $G$ is quantisable. Surprisingly, life becomes much easier when
considering line-graphs. Necessary and sufficient conditions for the
quantisability of $L^n(G)$ can actually be given and will be discussed
in the next section.

\subsection{Quantisable line-graphs}
We start by giving an answer to the question whether a given graph $H$
is the line-graph of another graph. The following are necessary and
sufficient conditions given by Richards \cite{Ri67}, see \cite{He78}
for a comprehensive overview over other equivalent statements.
\begin{theorem} \label{thldg}
  Let $H$ be a digraph and $A^{H}$ be its adjacency matrix. The
  following statements are equivalent:
  \begin{enumerate}
    \renewcommand{\theenumi}{(\roman{enumi})}
    \renewcommand{\labelenumi}{\theenumi}
  \item $H$ is a line-digraph;
  \item any two rows of $A^H$ are either identical or orthogonal;
  \item any two columns of $A^H$ are either identical or orthogonal.
  \end{enumerate}
\end{theorem}
It should be noted that a line-graph does in general not specify
uniquely its ancestor graph. This non-uniqueness is caused by sources
and sinks, (that is vertices with only outgoing or incoming edges) or
isolated vertices in the line-graph, see \cite{He78}. This problem is
less relevant for quantisable line-graphs as will be shown later, we
will therefore not consider it here further.

An immediate consequence of theorem~\ref{thldg} is the following
necessary and sufficient condition for a line-graph to be quantisable:
\begin{corollary} \label{codld}
  Let $H$ be a digraph with $N$ vertices and $A^{H}$ be its adjacency
  matrix. Then $H$ is a quantisable line-digraph iff there exist
  permutation matrices $P$ and $Q$ such that $PA^HQ$ is block-diagonal
  of the form
  \begin{equation} \label{eqdld}
    P A^H Q = \left( \begin{array}{cccc}
      J_{n_1} & & & \\
      & J_{n_2} & & \\
      & & \ldots & \\
      & & & J_{n_k} \\
      \end{array} \right)
  \end{equation}
  where $J_n$ is the square matrix of dimension $n$ containing only
  $1$'s and
  \begin{equation} \label{eqsbj}
    \sum_{i=1}^{k} n_i = N \, .
  \end{equation}
\end{corollary}
Proof: Follows directly from theorem~\ref{thldg}. The identical rows
(and columns) of $A^H$ form submatrices of $A^H$ containing only $1$'s. 
These submatrices have to be square matrices in order to have the same
pattern as a unitary matrix. The last condition (\ref{eqsbj}) follows
from the fact that a unitary matrix can not have a zero row or column.

The number of submatrices $k$ in (\ref{eqdld}) is equal to the number of 
vertices in
the ancestor graph and $n_i$ corresponds to the number of incoming and
outgoing edges at a vertex $v_i$ of the ancestor graph. The
corollary~\ref{codld} is thus equivalent to the statement
\begin{corollary} \label{coqlc}
  A graph $G$ has a quantisable line-graph $L(G)$ iff for every vertex
  $v_i$ in $V(G)$ the number of outgoing edges equals the number of
  incoming edges, that is, $d^+(v_i) = d^-(v_i)$.
\end{corollary}
The number of incoming and outgoing edges may of course vary from
vertex to vertex.

Quantisability of a line-graph turns out to be a rather strong
condition. Disregarding possible isolated vertices in the ancestor
graph, we can make the following statements about the ancestor graph
of a quantisable line-graph:
\begin{corollary} \label{coqle}
  Let $H$ be a quantisable line-graph and $G$ the ancestor of $H$;
  this implies
  \begin{enumerate}
    \renewcommand{\theenumi}{(\roman{enumi})}
    \renewcommand{\labelenumi}{\theenumi}
  \item the ancestor graph $G$ is uniquely defined by $H$ up to graph
    isomorphism;
  \item $G$ is either strongly connected or disconnected; if it is
    disconnected, then each of the disconnected components is strongly
    connected.
  \end{enumerate}
\end{corollary}

\subsection{Quantisable families of line-graphs}
We now turn to the question, whether a given graph $G$ has a
quantisable n-th generation line-graph $L^n(G)$. A necessary and
sufficient condition is given by the following theorem
\begin{theorem} \label{thnlq}
  A graph $G$ has a quantisable n-th generation line-graph $L^n(G)$
  iff for every vertex $v_i \in V(G)$ and every $v_j \in N^{(n-1)}_+
  (v_i)$, that is, for every vertex $v_j$ which can be reached from
  $v_i$ in $n-1$ steps, one finds
  \begin{equation}
    d^-(v_i) = d^+(v_j) \, .
  \end{equation}
\end{theorem}
Equivalently, one can write the condition above in terms of the
$(n-1)$-step in-neighbourhood $N^{(n-1)}_-$ of the vertices of $G$. \\
Proof: Let us start by giving the condition for the $(n+1)$-st
generation line-graph to be quantisable; from corollary~\ref{coqlc}
one obtains that $L^{n+1}(G)$ is quantisable, iff every vertex
$v^{(n)}_i \in V(L^n(G))$ has as many incoming as outgoing edges. We
may thus write
\begin{equation} \label{eqnlq}
  \sum_{j} A^{L^n(G)}_{ji} = \sum_{k} A^{L^n(G)}_{ik}
    \qquad {\mathrm{for\ all\ }} i,
\end{equation}
and the sum runs over all possible vertices of $L^n(G)$. A vertex
$v_i^{(n)} \in V(L^n(G))$ can be written in terms of $n$-step paths in
the original graph $G$, that is,
\begin{equation}
  v_i^{(n)} \equiv (v_{i_0}, v_{i_1}, \ldots, v_{i_n})
    \quad \mbox{for a set of vertices with} \quad
    A^G_{i_0i_1}\cdot A^G_{i_1i_2}\cdots A^G_{i_{n-1}i_n} \ne 0
\end{equation}
We now write eqn. (\ref{eqnlq}) in the form
\begin{equation}
  \sum_{i} A^{L^n(G)}_{ij} = \sum_{i_0} A^{L^n(G)}
    _{i_0 \ldots i_n, i_1 \ldots i_{n+1}} = A^G_{i_1i_2} \cdot
    A^G_{i_2i_3} \cdot \ldots \cdot A^G_{i_ni_{n+1}} \sum_{i_0} A^G_{i_0i_1}
\end{equation}
and
\begin{equation}
  \sum_{i} A^{L^n(G)}_{ji} =
    \sum_{i_{n+2}} A^{L^n(G)}_{i_1 \ldots i_{n+1}, i_2 \ldots i_{n+2}}
    = A^G_{i_1i_2} \cdot A^G_{i_2i_3} \cdot \ldots \cdot A^G_{i_{n}i_{n+1}}
    \sum_{i_{n+2}} A^G_{i_{n+1}i_{n+2}} \, .
\end{equation}
We thus obtain the condition
\begin{equation}
  d^-(v_{i_1}) = \sum_{i=1}^N A^G_{i i_1} = \sum_{i=1}^N A^G_{i_{n+1}i}
    = d^+(v_{i_{n+1}}) \quad \mbox{if} \quad A^G_{i_1i_2} \cdot
    A^G_{i_2i_3} \cdot \ldots \cdot A^G_{i_ni_{n+1}} \ne 0 \, ,
\end{equation}
that is, if there exists a path to reach $v_{i_{n+1}}$ from
$v_{i_{1}}$ in $n$ steps; this completes the proof of the theorem.

Equivalently, theorem~\ref{thnlq} may be expressed as
\begin{corollary}
  A graph $G$ with adjacency matrix $A^G$ has a quantisable
  n-th generation line-graph $L^n(G)$ iff for every pair of vertices
  $v_i, v_j \in V(G)$
  \begin{equation}
    d^-(v_i) = d^+(v_j) \quad \mbox{whenever} \quad
    \left( (A^G)^{n-1} \right)_{ij} \ne 0 \, ,
  \end{equation}
  where $(A^G)^{n-1}$ denotes the $(n-1)$st power of the matrix $A^G$.
\end{corollary}

It is clear from the conditions above that more and more restrictions
are imposed on $G$ if one wants to construct families of line-graphs
with an increasing number of quantisable line-graph generations. In
the following we give a couple of general statements on line-graph
families. We assume here that the ancestor graph $G$ is connected; a
generalisation to disconnected line-graphs is obvious in the light of
corollary~\ref{coqle}.
\begin{corollary} \label{coqlf}
  Let $G$ be a digraph and $L^n(G)$ its family of line-graphs;
  \begin{enumerate}
    \renewcommand{\theenumi}{(\roman{enumi})}
    \renewcommand{\labelenumi}{\theenumi}
  \item
    $L^n(G)$ is quantisable for all $n$ iff $G$ is regular, that is,
    iff for every pair of vertices $v_i, v_j \in V(G)$, $d^+(v_i) =
    d^+(v_j) = d^-(v_i)= d^-(v_j)$.
  \item \label{itfap}
    Let $G$ be a graph with a primitive adjacency matrix $A^{G}$, that
    is, there exists an integer $k$ such that $(A^{G})^k$ has all
    matrix elements strictly positive. Then $L^n(G)$ is quantisable
    for $n \ge k$ iff $G$ is regular.
  \item \label{itfbg}
    A graph of order $N$ is called bipartite, denoted $K_{N_1,N_2}$,
    if there exist two distinct sets of vertices $V_1$ and $V_2$ with
    $N_1$ and $N_2$ elements, respectively, $N_1 + N_2 = N$, such that
    every vertex in $V_1$ is connected to every vertex in $V_2$ but
    not to any vertex in $V_1$ and {\it vice versa}. For $N_1 \ne N_2$
    we have: A line-graph $L^n(K_{N_1,N_2})$ is quantisable iff $n$ is
    an odd integer.
  \item
    An $r$-partite graph $K_{N_1, N_2, \ldots, N_r}$ is defined in
    analogy to a bipartite graph. For $r\ge 3$ and whenever at least
    two of the $r$ vertex sets contain a different number of vertices
    one obtains: the line-graph generations are quantisable for $n=1$
    only. 
  \item
    The first generation line-graph of an undirected graph without
    isolated vertices is quantisable.
  \end{enumerate}
\end{corollary}
The above list is only a small selection of possible conclusions
following directly from theorem~\ref{thnlq} for some important classes
of graphs. Many more could be formulated here. It becomes clear from
the examples that quantisability is a very restrictive condition.
Especially point~\ref{itfap} in corollary~\ref{coqlf} is important in
connection with corollary~\ref{coqle}. Strongly connected graphs are
typically primitive; only graphs with additional structure like
bipartite graphs do not fall into this class. Line-graph families with
infinitely many members being quantisable thus implies a high degree
of regularity in the graph.

In the next section we will discuss some examples of quantisable
line-graph families and study the spectral properties of matrices of 
the associated unitary stochastic ensembles.

\section{Examples} \label{secex}
\subsection{De Bruijn graph families}
The aim of this section is to present the statistical properties of
ensembles of unitary matrices corresponding to quantisable line-graph
families of regular initial graphs $G$. One set of such families
consists of de Bruijn graphs of $M$-th order \cite{BJ01}. They are
obtained as the line-graph families of fully connected initial
digraphs $K_M$ with $A^{K_M} = J_M$, that is, 
\begin{equation}
  V(K_M) = \{1,\ldots,M\}, \qquad E(K_M) = \{(ij): i,j \in V(K_M)\}.
\end{equation}
The graphs $K_M$ have $M$ vertices and $M^2$ bonds connecting each
vertex with all other including itself, so they have $M$ loops.
The line-graph families $L^n(K_M)$ have accordingly $M^{n+1}$ vertices
and $M^{n+2}$ edges with $M$ incoming and outgoing edges at each
vertex, that is, the $L^n(K_M)$ are all $M$-regular. The family
$L^n(K_2)$ are the family of binary graphs studied in \cite{Ta00}.
\begin{figure}[hbt]
  \begin{center} \includegraphics{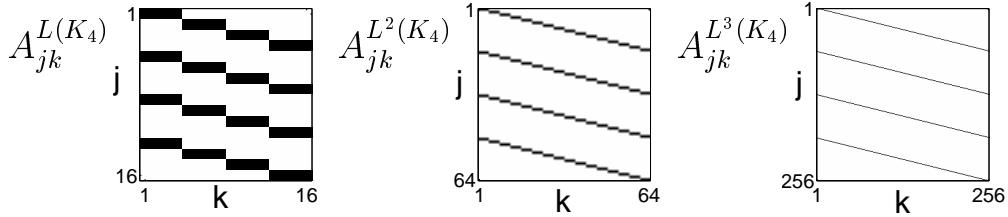} \end{center}
  \caption{Adjacency matrices $A^{L^n(K_4)}$ of 4th-order de Bruijn
    graphs generated as the line-graphs of the fully connected graph
    $K_4$, for $n=1$, $n=2$ and $n=3$. Nonzero entries are denoted as
    black squares. The matrix size equals 16, 64 and 256,
    respectively.}
  \label{fimbf}
\end{figure}

In the following, we will consider stochastic transition matrices
$T^{K_M}$ on the initial graph $K_M$ with constant transition
probabilities $1/M$ between all vertices, that is, $T^{K_M} =
\frac{1}{M} J_M$. These matrices saturate the well known van der
Waerden inequality concerning permanents of bistochastic matrices,
i.e. $\mbox{per}(T) \ge M!/M^M$ \cite{Ma79}. It is easy to see that
the $T^{L^n(K_M)}$ are unistochastic, since related unitary matrices
(\ref{eqqtc}) may be constructed out of discrete Fourier transforms of
size $M$, ${\mathcal{F}}_{ml}^{(M)} = \frac{1}{\sqrt{M}} \, e^{2\pi
i\, ml/M}$. The graphs $L^n(K_M)$ have topological entropy equal to
$\ln M$. The metric entropy of the process defined by $T^{L^n(K_M)}$
is also equal to $\ln M$.
\begin{figure}[hbt]
  \begin{center} \includegraphics{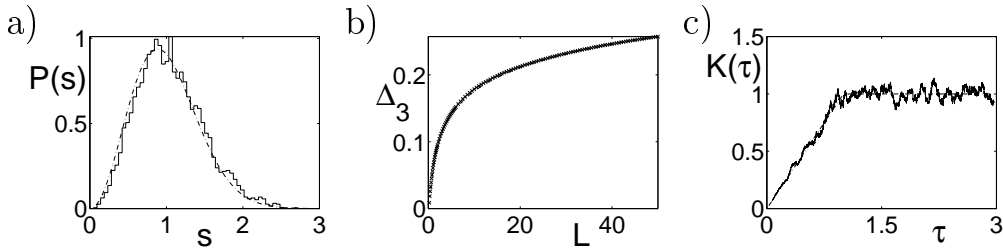} \end{center}
  \caption{Spectral statistics of a single unitary matrix of size
    $N=4096$ corresponding to the de Bruijn graph $L^5(K_4)$:
    (a) level spacing distribution $P(s)$, (b) spectral rigidity
    $\Delta_3(L)$ and (c) spectral form factor $K(\tau)$ (with
    $\Delta\tau=0.07$). CUE predictions (coinciding with numerical
    data in panel (b)) are represented by dot-dashed lines.}
  \label{fisbf}
\end{figure}

The adjacency (and transition) matrices for the stochastic process on
de Bruijn graphs represent a discrete generalization of the Bernoulli
shift. Three matrices from the family $L^n(K_4)$ are depicted in
Fig.~\ref{fimbf}. The non-zero elements are marked as black squares,
they are placed along four lines. In the limit of large $n$ the
structure of the matrices approaches the graph of the function
$4x|_{\rm mod 1}$ (Renyi map) rotated clockwise by angle $\pi/2$. Such
a correspondence between digraphs and classical dynamical systems has
been recently pointed out in \cite{Pa01}.
 
We are interested in the spectral properties of a generic quantum
propagator $U^{L^n(K_M)}$ corresponding to the Markov process on a de
Bruijn graph. By means of the discrete Fourier transform
${\mathcal{F}}$ we constructed a unitary propagator associated with
the stochastic transition matrix $T^{K_M}$. By multiplying with 
random diagonal unitary matrices $D_1$ and $D_2$ we obtain a typical
element $\tilde U$ of the ensemble (\ref{equse}). Fig.~\ref{fisbf}
shows the spectral statistics received from eigenphases of a single
unitary matrix of size $N=4096$ from the ensemble, $U^{L^5(K_4)}$.
The level spacing distribution $P(s)$, the spectral rigidity
$\Delta_3(L)$ \cite{Dy62} and the spectral form factor $K(\tau)$ (the
Fourier transform of the two point correlation function) \cite{Ha00}
are plotted. The statistics coincides well with the predictions of
random matrices for the  Circular Unitary Ensemble (CUE) \cite{Me91},
although it is only the fifth iteration of the line-graph
construction. The spectral form factor $K(\tau)$ was averaged over a
parameter window $\Delta\tau$. We have also obtained qualitatively
similar results averaging $K(\tau)$ over a unitary stochastic ensemble
as defined in (\ref{equse}) consisting of $10^3$ unitary matrices 
$\tilde U$ of size 64.

\begin{figure}[hbt]
  \begin{center} \includegraphics{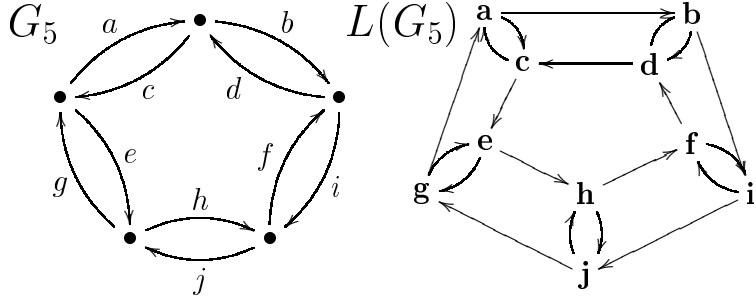} \end{center}
  \caption{Cycle graph family: $G_5$ and its line-graph $L(G_5)$.}
  \label{figcf}
\end{figure}

\subsection{Symmetric cycle graph family} \label{secsc}
\begin{figure}[hbt]
  \begin{center} \includegraphics{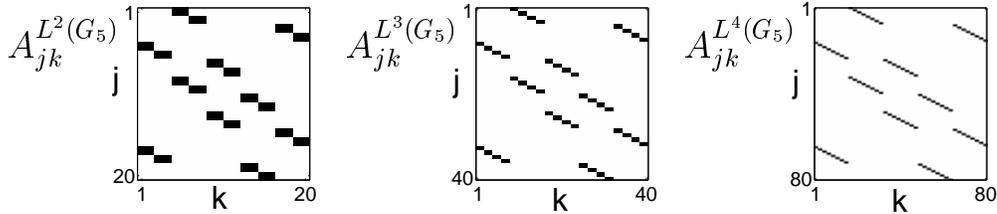} \end{center}
  \caption{Structure of the adjacency matrices of line-graphs family
    $L^n(G_5)$ obtained from the symmetric cycle graph $G_5$ with
    $n=2$, $n=3$ and $n=4$ with matrix size $N=20$, $40$, $80$,
    respectively.}
  \label{fimcf}
\end{figure}

Next we consider a family of quantisable line-graphs which are
constructed from symmetric cycle digraphs. A $M$-symmetric cycle graph
$G_M$ is an undirected graph with $M$ vertices placed on a circle each
vertex connected with its two neighbors only see Fig.~\ref{figcf}.
More formally,
\begin{equation}
  V(G_M)=\{1,\ldots,M\} \, , \qquad
  E(G_M)=\{(1M),(M1)\} \cup \{(i\,\,i+1),(i+1\,\,i):i=1\ldots M-1\} \, .
\end{equation}
The initial digraph $G_M$ is a 2-regular graph which implies that its
line-graphs $L^n(G_M)$ are all quantisable following
corollary~\ref{coqlf}. The n-th line-graph generation has $M \cdot
2^n$ vertices, see Fig.~\ref{figcf}. Next, we choose a stochastic
process with equal probabilities, $1/2$, to move from a given vertex
to one of its neighbors. The topological and metric entropies are both
equal to $\ln 2$ in this case. The non-zero matrix elements of the
adjacency matrices in the family have the same structure for any fixed
$M$, see Fig.~\ref{fimcf} for the family $L^n(G_5)$.
Cycle graphs and their quantisation play an important role in the
study of Anderson-type localisation in one-dimensional diffusive
systems. A full description of localisation in terms of return
probabilities on infinite chains has been given by Schanz and
Smilansky \cite{Sc00}, cycle graphs have also been discussed in
\cite{Ta01} in connection with the spectral gap of the corresponding
Markov process. One finds
\begin{equation}
  \Delta_{T^{G_M}} \sim M^{-2}
\end{equation}
that is, the spectral gap vanishes for large $M$. We may now consider
two limits: by fixing the generation $n$ of the line-graph $L^n(G_M)$
and looking at the limit $M \rightarrow \infty$ one indeed finds
deviation from RMT due to localisation \cite{Ta01}; we may on the
other hand fix $M$ and increase $n$ which produces line-graphs with an
increasing number of vertices but constant spectral gap and we expect
RMT-behaviour in this limit.
\begin{figure}[hbt]
  \begin{center} \includegraphics{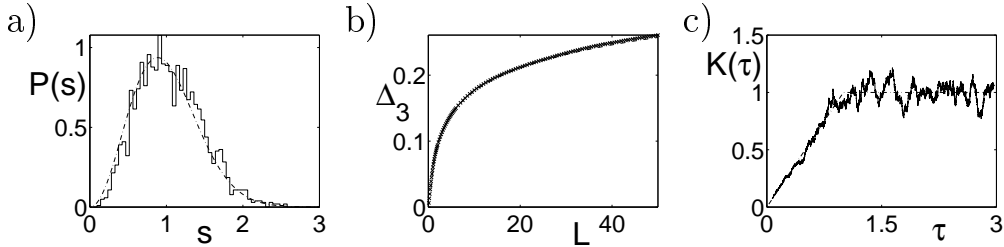} \end{center}
  \caption{As in Fig.~\ref{fisbf} for a single random unitary matrix
    of size $5120$ associated with the digraph $L^{10}(G_5)$.}
  \label{fiscf}
\end{figure}

This is indeed what is observed; in Fig.~\ref{fiscf} the statistics
obtained from a quantum propagator $U^{L^{10}(G_5)}$, with randomly
chosen phases conforms well with the prediction of CUE.
\begin{figure}[hbt]
  \begin{center} \includegraphics{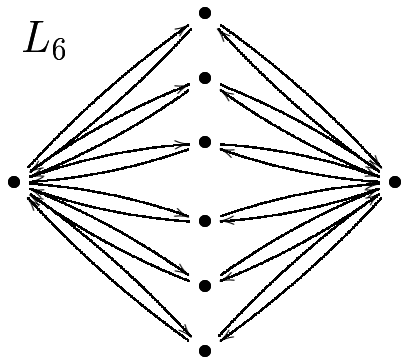} \end{center}
  \caption{Bipartite digraph $K_{2,6}$ --- the initial digraph in the
    bipartite graph family $L^n(K_{2,6})$.}
  \label{fig2f}
\end{figure}

\subsection{Bipartite graph family $L^n(K_{2,M})$}
As a last example we will have a look at bipartite digraphs
$K_{N_1,N_2}$, see corollary~\ref{coqlf}\ref{itfbg}. In terms of its
vertex and edge set, $K_{N_1,N_2}$ is defined as
\begin{equation}
  V( K_{N_1,N_2} ) = V_1 \cap V_2 \, , \ \ V_i = \{1,\ldots ,N_i\} ,
    \quad E( K_{N_1,N_2} ) = \{(ij),(ji): i \in V_1, j \in V_2\} .
\end{equation}
\begin{figure}[hbt]
  \begin{center} \includegraphics{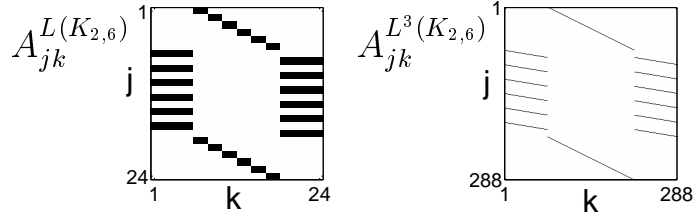} \end{center}
  \caption{Structure of the adjacency matrices for line-graphs of
    bipartite graphs $L^n(K_{2,6})$ for $n=1$ and $n=3$; the
    corresponding matrix sizes are $N=24$ and $N=288$.}
  \label{fim2f}
\end{figure}

\noindent
A class of bipartite graphs, which has been studied recently in the
context of spectral statistics of quantum graphs are so-called star
graphs $K_{1,M}$ which have one central vertex and $M$ arms \cite{Ko99,
Be01, Ta01}. Using the quantisation condition employed by Kottos and
Smilansky \cite{Ko99} leads to quantum propagation with an associate
transition matrix with spectral gap scaling like $\Delta_M \sim 1/M$;
again one finds deviations of the spectral statistics from RMT
persisting in the large $M$ limit. By fixing $M$ and considering the
line-graph family $L^n(K_{1,M})$, which is quantisable for $n$ odd, see
corollary~\ref{coqlf}, we can again achieve a large matrix limit with
non-vanishing spectral gap. We indeed find convergence to the RMT -
statistics to a degree very similar to Figs.~\ref{fisbf} and~\ref{fiscf}.
We also studied bipartite graphs $K_{2,M}$ and its line-graphs. An
example of such a graph is plotted in Fig.~\ref{fig2f}. The graphs in
the family $L^n(K_{2,M})$ have vertices with either 2 or $M$ outgoing
edges and are again quantisable for $n$ odd. A unistochastic
transition matrix may be constructed choosing probabilities $1/2$ for
vertices with 2 outgoing bonds and $1/M$ otherwise. Figure~\ref{fim2f}
shows the structure of the non-zero elements of transition matrices
for $L(K_{2,6})$ and $L^3(K_{2,6})$. 
 
The construction of a unitary quantum map may be achieved by means of
the discrete Fourier transform. As for the previous examples, the
spectral statistics of eigenphases of $U^{L^n(K_{2,6})}$ follows CUE to
the same degree as shown in Figs.~\ref{fisbf} and~\ref{fiscf} for $n
\ge 5$. 
 
\section{Conclusions} \label{secco}
By constructing directed line-graphs from an arbitrary initial digraph
$G$ one obtains a family of graphs with in general increasing number
of vertices but identical topological and metric properties. We showed
that all digraphs in such a family indeed have the same set of
periodic orbits and that furthermore the non-zero eigenvalues of the
adjacency matrices of graphs from the same family are identical. Next
we considered stochastic Markov processes on a digraph and defined the
corresponding process on its line-graph. We demonstrated that both
processes have the same metric entropy and the transition matrices
describing the processes have the same non-zero eigenvalues. The
construction of the line-graph family is in fact a method to translate
a finite Markov processes to a larger space preserving its topological
and metric properties.  

We gave necessary and sufficient conditions for a line-graph to be
quantisable and gave examples of line-graph families $L^n(G)$ which
can be quantised for infinitely many $n$. The line-graph construction
thus makes it possible to consider a semiclassical limit of large
matrix size for unitary ensembles on graphs with fixed `classical',
i.e. stochastic dynamics. This method complements an idea developed in
a previous paper \cite{Pa01}, in which a semiclassical limit was
considered by looking for a specific dynamical system associated with
an initial graph.
 
We would like to stress again that the problem of finding necessary
and sufficient conditions for a general bistochastic matrix $T$ to be
unistochastic is still open \cite{Zy01}. Such conditions can, however,
be given for line-graphs and turn out to be very restrictive. One way
to enlarge the number of graph families with well defined classical
limit is to consider unitary matrices $U^n$ and associated transition
matrices $T^{(n)}$ for which topological and metric properties
converge to fixed values in the limit of large matrix sizes
\cite{Pa02}.

Quantum maps generated from Markov processes on 
families of line-graphs considered here all display CUE statistics in
their spectral fluctuations. These results were observed for families
originating from fully connected digraphs (de Bruijn), symmetric
cycles and bipartite digraphs of the form $K_{1,M}$ and $K_{2,M}$. This
behaviour is attributed to the fact that the spectral gap is positive
and constant under the line-graph iteration in all cases whereas the
number of vertices increases with $n$. The results thus confirm the
conjecture~\ref{cormt}, which relates the size of the spectral gap of
the classical transition matrix and the spectral statistics of the
associated ensemble of random unitary matrices.
 
\section*{Acknowledgement}
We would like to thank Wojciech S{\l}omczy\'nski and Simone Severini
for stimulating discussions and helpful remarks. Financial support by
Polish Komitet Bada\'n Naukowych under the grants no 2~P03B~072~19 and
no 5~P03B~086~21 as well as by the Royal Society is gratefully
acknowledged.

\end{document}